\documentclass[letter]{emulateapj}
\usepackage{natbib}
\usepackage{epsfig}

\usepackage{amssymb}
\usepackage{latexsym}
\usepackage{ifthen}
\usepackage{url}

\newcommand{\degrees}{\,^\circ}
\newcommand{\msun}{\,M_{\odot}}
\newcommand{\tsun}{T_{\odot}}

\newcommand{\pb}{P_b}
\newcommand{\pbdot}{\dot{P}_b}

\newcommand{\fhour}{^{\mathrm{h}}}
\newcommand{\fmin}{^{\mathrm{m}}}
\newcommand{\fsec}{\mbox{\ensuremath{.\!\!^{\mathrm{s}}}}}
\newcommand{\fdeg}{^{\circ}}

\newcommand{\psr}[1]
{\ifthenelse{\equal{#1}{0737}}{PSR~J0737$-$3039}{\ifthenelse{\equal{#1}{1534}}{PSR~B1534+12}{\ifthenelse{\equal{#1}{1913}}{PSR~B1913+16}{\ifthenelse{\equal{#1}{1802}}{PSR~J1802$-$2124}{\ifthenelse{\equal{#1}{1756}}{PSR~J1756$-$2251}{\ifthenelse{\equal{#1}{1906}}{PSR~J1906+0746}{\ifthenelse{\equal{#1}{1141}}{PSR~J1141$-$6545}{\ifthenelse{\equal{#1}{2127}}{PSR~B2127+11C}{\ifthenelse{\equal{#1}{0751}}{PSR~J0751+1807}{\ifthenelse{\equal{#1}{1713}}{PSR~J1713+0747}{\bf ???????}}}}}}}}}}}

\slugcomment{Draft \today; accepted for publication in the Astrophysical Journal}
\shorttitle{Mass measurement of PSR~J1802$-$2124}
\shortauthors{Ferdman et al.}
\begin{document}

\title{A precise mass measurement of the intermediate-mass binary pulsar \psr{1802}}

\author{
R. D. Ferdman\altaffilmark{1,2,3}, 
I. H. Stairs\altaffilmark{3,4,5}, 
M. Kramer\altaffilmark{6,7}, 
M. A. McLaughlin\altaffilmark{8,9}, 
D. R. Lorimer\altaffilmark{8,9}, 
D. J. Nice\altaffilmark{10}, 
R. N. Manchester\altaffilmark{4}, 
G. Hobbs\altaffilmark{4}, 
A. G. Lyne\altaffilmark{7},
F. Camilo\altaffilmark{11}, 
A. Possenti\altaffilmark{12},
P. B. Demorest\altaffilmark{13},
I. Cognard\altaffilmark{1,2},
G. Desvignes\altaffilmark{1,2},
G. Theureau\altaffilmark{1,2,14},
A. Faulkner\altaffilmark{7} and
D. C. Backer\altaffilmark{15} 
}

\altaffiltext{1}{Station de Radioastronomie de Nan\c{c}ay, Observatoire de Paris, 18330 Nan\c{c}ay, France; robert.ferdman@obs-nancay.fr}
\altaffiltext{2}{Laboratoire de Physique et Chimie de l'Environnement et de l'Espace, Centre National de la Recherche Scientifique, F-45071 Orl\'{e}ans, Cedex 2, France}
\altaffiltext{3}{Department of Physics and Astronomy, University of British 
  Columbia, Vancouver, BC V6T 1Z1, Canada}
\altaffiltext{4}{Australia Telescope National Facility, CSIRO, Epping NSW 1710, Australia}
\altaffiltext{5}{Centre for Astrophysics \& Supercomputing, Swinburne University of Technology, Mail 39 PO Box 218 Hawthorn Vic 3122 Australia}
\altaffiltext{6}{Max-Planck-Institut f\"{u}r Radioastronomie, Auf dem H\"{u}gel 69, 53121, Bonn, Germany}
\altaffiltext{7}{University of Manchester, Jodrell Bank Centre for Astrophysics, Alan Turing Building, Oxford Road, Manchester M13 9PL, United Kingdom}
\altaffiltext{8}{Department of Physics, West Virginia University, Morgantown, WV 26505} 
\altaffiltext{9}{National Radio Astronomy Observatory, Green Bank, WV 24944} 
\altaffiltext{10}{Physics Department, Bryn Mawr College, Bryn Mawr, PA 19010}
\altaffiltext{11}{Columbia Astrophysics Laboratory, Columbia University, New York, NY 10027}
\altaffiltext{12}{INAF - Osservatorio Astronomico di Cagliari, Loc.~Poggio dei Pini, 09012 Capoterra (CA), Italy}
\altaffiltext{13}{National Radio Astronomy Observatory, Charlottesville, VA 22901}
\altaffiltext{14}{GEPI, Observatoire de Paris, Centre National de la Recherche Scientifique, Universit\'{e} Paris Diderot, 92195 Meudon, France}
\altaffiltext{15}{Department of Astronomy and Radio Astronomy Laboratory, University of California, Berkeley, CA 94720}

\begin{abstract}
PSR~J1802$-$2124 is a 12.6-ms pulsar in a 16.8-hour binary orbit with a relatively massive white dwarf (WD) companion.  These properties make it a member of the intermediate-mass class of binary pulsar (IMBP) systems.  We have been timing this pulsar since its discovery in 2002.  Concentrated observations at the Green Bank Telescope, augmented with data from the Parkes and Nan\c{c}ay observatories, have allowed us to determine the general relativistic Shapiro delay.  This has yielded  pulsar and white dwarf mass measurements of $1.24\pm0.11\msun$ and $0.78\pm0.04\msun$ ($68\%$ confidence), respectively.  The low mass of the pulsar, the high mass of the WD companion, the short orbital period, and the pulsar spin period may be explained by the system having gone through a common-envelope phase in its evolution.  We argue that selection effects may contribute to the relatively small number of known IMBPs. 
\end{abstract}


\section{Introduction}
\setcounter{footnote}{0}
In the standard theory of pulsar spin-up, a neutron star (NS) in a binary system accretes matter from its companion star.  This serves to transfer angular momentum to the NS, increasing the spin frequency of the pulsar \citep[e.g.,][]{acrs82}.  The type and duration of mass transfer onto the pulsar determines the final spin period and depends a great deal on the nature and evolution of the system 
\citep[for reviews of binary pulsar systems and their evolution; e.g.,][]{bv91,pk94,sta04a,tv06corr}.  
 
The measured mass distribution of NSs in pulsar binary systems is more diverse than previously thought.  Observations show that many pulsars have masses which lie significantly outside the oft-cited statistical average of $1.35\pm0.04\,\msun$ \citep{tc99}.  This highlights the need to invoke a variety of evolutionary scenarios in order to explain the collection of observed pulsar binary systems.  Most double-neutron-star (DNS) binary systems are thought to have undergone common-envelope (CE) evolution, in which the neutron star resulting from the first supernova spirals into the envelope of the companion star; the CE is subsequently expelled from the system \citep[e.g.,][]{tv06corr}.  That this mass-transfer stage in the evolution of these systems is short-lived is evidenced in part by the relatively long $\sim 10-100$-ms pulsar spin periods observed in DNS binaries.

The majority of known neutron star-white dwarf (NS-WD) binaries have pulsars that spin with millisecond periods ($<10$\,ms) and have very low eccentricities.  Such short rotation periods indicate that the pulsars have undergone a relatively long, stable period of accretion of material from the companions' outer envelopes, during which they are seen as low-mass X-ray binaries \citep{bv91,wv98,asr+09}.  In the process, the matter-donating stars lose an appreciable amount of mass.  This conclusion is supported by the relatively low masses of the WDs found in these binaries, referred to as low-mass binary pulsar (LMBP) systems \citep[e.g.,][]{tv06corr}.

In contrast to the LMBP binaries are the intermediate-mass binary pulsar class \citep[IMBPs;][]{cnst96,eb01a}.  These systems are characterized by pulsar spin periods of tens of milliseconds, massive carbon-oxygen (CO) or oxygen-neon-magnesium (O-Ne-Mg) WD companions ($\gtrsim 0.4\,\msun$), and orbital eccentricities which, while still small, are significantly larger than those of LMBP systems.

\begin{deluxetable*}{llccccc}
\tablecaption{Summary of observations of PSR J1802$-$2124\label{tab:obs_1802}}
\tablecolumns{7}
\tablehead{
  \colhead{Telescope}  &  
  \colhead{Instrument}  & 
  \colhead{Center}  &  
  \colhead{Total effective}  &   
  \colhead{Integration} &
  \colhead{Number} &
  \colhead{Date span}  \\
  \colhead{}  &   
  \colhead{}  & 
  \colhead{frequency}  &  
  \colhead{bandwidth}  &  
  \colhead{time} &
  \colhead{of TOAs} &
  \colhead{(MJD)} \\
  \colhead{}  &   
  \colhead{}  & 
  \colhead{(MHz)}  &  
  \colhead{(MHz)}  &  
  \colhead{(min)} &
  \colhead{} &
  \colhead{}
}
\startdata
 Parkes       & Filterbank  &  1390   &  256 &  typically 20  &  \phn106 & 52605-54910 \\
  GBT         & GASP        &  1400   &  $64-96$   &  \phn3       & 2233 &  53441-54950 \\
  Nan\c{c}ay  & BON         &  1398   &  $64-128$  &  14      & \phn107 &  54188-54806
\enddata
\end{deluxetable*}

Several formation scenarios have been suggested to explain the existence of IMBPs \citep[e.g.,][]{li02corr}.  One idea is that, as with DNS binaries, the NS spirals into the envelope of its companion to form a CE which is then promptly ejected from the system \citep[e.g.,][]{vdh94}.  This is supported by the short orbital periods ($\pb$) seen in many of the IMBP systems.  It has also been proposed, however, that a neutron star within the envelope of its companion will be forced to undergo hypercritical accretion, becoming a black hole and thus rendering the system unobservable \citep[e.g.,][hereafter B01]{che93,bro95,blpb01}.  

\citet{tvs00} argued that systems with heavy CO WD companions and orbits with $3 \lesssim \pb \lesssim 70$ days can undergo, and survive, a short-lived phase of highly super-Eddington mass transfer to the NS.  Here, the inspiral that results in a CE is avoided if the re-radiated accretion energy is great enough to evaporate most of the transferred material before it approaches the NS too closely \citep[see also][]{tkr00}.  Still, this scheme does not work for IMBPs with $\pb \lesssim 3$ days, suggesting the need to invoke CE evolution to explain their existence.  B01 put forward a possible alternate formation scenario in which the two progenitors are main sequence stars of similar mass, which evolve to form two helium cores.  This is similar to a related scenario for the evolution of close DNS binaries \citep{bro95}, the difference being that to form an IMBP, one of the stars would be just below the mass threshold for NS formation, becoming a WD instead.  It is clear that the evolution of IMBPs remains an open question. 

There are now sixteen known IMBP systems.  Until now, only one of these, PSR~J0621+1002, has a pulsar with a measured mass \citep[$m_1=1.70^{+0.10}_{-0.16}\,\msun$;][]{nsk08}, which is somewhat larger than the range of masses seen in well-measured DNS systems \citep[e.g.,][]{sta08}.  Here we present results from timing of \psr{1802}, another member of the class of IMBPs, discovered in the Parkes Multibeam Pulsar Survey \citep{fsk+04}.  In the case of \psr{1802}, its heavy companion led us to believe that the system would be a good candidate for measuring the Shapiro delay of the pulses in the gravitational potential of the white dwarf.  We were, in fact, able to measure such an effect, enabling us to determine the individual masses of each member of the binary system.  In this article, we discuss these measurements and their implications for reconstructing the formation and evolutionary histories of this system and others like it\footnote{A preliminary version of these results was reported in \citet{fer08}}.

\section{Observations}

We used three observatories to collect pulsar data. In what follows, we describe the instruments used and the data obtained at each of these telescopes. A summary of the observing details is found in Table~\ref{tab:obs_1802}.

\subsection{Parkes}
We have added to the data set presented by \citet{fsk+04} using the 64-m Parkes telescope in Australia.  Observations were carried out at regular intervals using a $2 \times 512 \times 0.5$-MHz filterbank centered at 1390\,MHz, each typically 20 minutes in duration.  The data from each channel were detected and the two polarizations summed in hardware before 1-bit digitization every $80-250\,\mu$s.  The data were recorded to tape and subsequently folded off-line.  Parkes data used for this work were collected at 106 epochs over 6.3 years.  This long timing baseline was particularly useful for measurements of astrometric parameters.  Discovery and some initial timing data were taken with a different filterbank (3-MHz channels) and were excluded from our analysis.

\subsection{Green Bank}
We also used the 100-m Robert C.~Byrd Green Bank Telescope (GBT) in West Virginia.  Data-taking at the GBT was performed with the Green Bank Astronomical Signal Processor \citep[GASP;][]{dem07}.  GASP is a flexible baseband system, which performs 8-bit Nyquist sampling of the incoming data stream at $0.25 \mu$s intervals in both orthogonal polarizations.  The signal was divided into $16$ or $24 \times 4$-MHz channels\footnote{The number of channels used occasionally varied due to radio frequency interference and available computing resources.} centered near 1400\,MHz.  The incoming data stream was then coherently dedispersed \citep{hr75} in software.  After this, the signals were folded at the pulse period to form pulse profiles, typically representing 3-minute integration times.  These were usually flux-calibrated in each polarization by using as a reference the signal from a noise diode source that was injected at the receiver.  When calibration data were not available, we normalized the profile data in each polarization by the root-mean-square (rms) value of the corresponding off-pulse signal.  The data were finally summed over both polarizations and across all frequency channels to give the total power signal \citep[for further details on GASP operation and data reduction, see][]{dem07,fer08}.   

Using the GBT, we obtained a total of 42 epochs of data that span more than 4 years.  Ten of these observations consisted of observing sessions between 6 and 8 hours long. These were scheduled so as to sample as fully as possible the orbit of the system, of particular importance for the detection of Shapiro delay.  The rest of the data consisted of approximately 1-2 hours of observing.  These GBT observations comprised the vast majority of our data set ($65\%$ by time, $97\%$ by weight). 

\subsection{Nan\c{c}ay}

We included in our data set observations of \psr{1802} taken by the 94-m circular-dish equivalent Nan\c{c}ay telescope in France.  These data were recorded with the Berkeley-Orl\'{e}ans-Nan\c{c}ay (BON) pulsar backend, a sister system to the GASP instrument at the GBT.  The BON instrument is also a baseband recorder, which performs coherent dedispersion on the incoming data stream in real time.  The data originally consisted of $16 \times 4$-MHz channels, and since 2008 July 25 has been increased to include 32 frequency channels, centered at 1398\,MHz in both cases.  As with the GASP backend, the data were detected and folded after signal dedispersion was performed.  Flux calibration was not available for the Nan\c{c}ay data, and so we normalized each hand of polarization by its off-pulse rms signal before obtaining total power pulse profiles.  This is acceptable since Nan\c{c}ay is a meridian-style telescope, and observed \psr{1802} for approximately 1 hour per day; during this time, the telescope gain was not expected to change significantly.  In all, we collected data at 26 epochs over 1.7 years.  The output profiles represent summation across the observing bandwidth with a typical integration time of 14 minutes, or approximately 4 scans per observing day.  Although the Nan\c{c}ay data set is small compared with that of the GBT data, it proved useful in filling an observational gap between 2007 April 18 and 2007 October 2007, during which time the GBT was undergoing track repair.

\section{Timing Analysis}

\begin{figure}[t]
\epsscale{1.}
\plotone{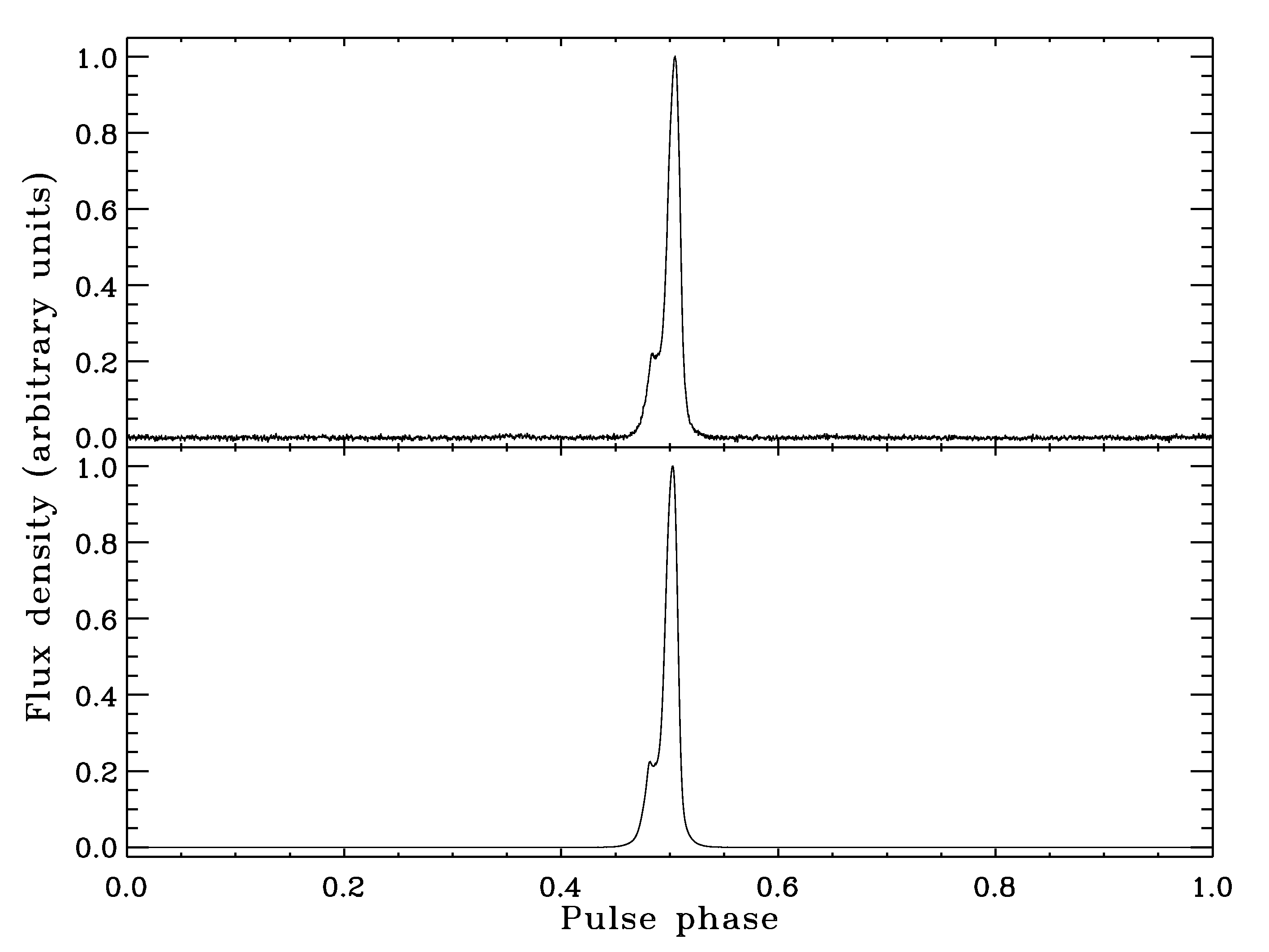}
\caption{Template profile for PSR~J1802$-$2124. {\it Top}: High signal-to-noise profile created by all useable data taken at the GBT with the GASP pulsar backend instrument. {\it Bottom}:  Profile created by fitting multiple Gaussians to the PSR~J1802$-$2124 pulse profile.
\label{fig:std_prof}}
\end{figure}
In order to determine the pulse times-of-arrival (TOAs), we constructed a template profile for \psr{1802}. This was done by first aligning in phase all individual GBT-derived 1400-MHz pulse profiles, then computing a simple summation of the data, weighted by the noise in the off-pulse regions of the input profiles.  Finally, we performed a multiple-Gaussian fit to this high signal-to-noise profile, obtaining a zero-noise reference template \citep{kwj+94,kra94}, shown in Figure \ref{fig:std_prof}.  This method allowed us to eliminate low-level ($\lesssim1\%$ of the peak height) structure that was seen to either side of the main pulse in the cumulative profile; we suspect that this was likely due to aliasing that occurs when the channel filters do not cut off sharply at the Nyquist frequency \citep[see, e.g.,][]{shr97}.  This noise-free template was then used to calculate TOAs from both GBT and Nan\c{c}ay data, as their hardware configuration and output profile data format are similar.  The same template was also applied to the Parkes data, but was first computed with 256 bins, then convolved with a top-hat function representing the in-channel dispersive smearing at 1390 MHz, then reduced to 64 bins.  This was done using the {\tt psrchive} software \citep{hvm04}.

Pulse TOAs were then calculated by cross-correlating each pulse profile with the reference template profile in the frequency domain \citep{tay92}.  The time offset corresponding to each of the phase shifts found in the correlation was then added to the time-stamp recorded for each profile, resulting in a TOA that represents a time very close to the midpoint of each particular integration.  In total, we measured 2446 TOAs: 106 from Parkes data, 2233 from GBT data, and 107 from Nan\c{c}ay data.

A model ephemeris was then fitted to the topocentric TOAs, using the \texttt{tempo} software package\footnote{\texttt{http://www.atnf.csiro.au/research/pulsar/tempo}}.  Included in this model is the motion of the Earth, calculated using the JPL DE405 Solar System model \citep{sta04b}.  To account for any instrumental and standard template profile differences, we fit for arbitrary time offsets for the Parkes and Nan\c{c}ay-derived TOAs, with reference to those from the GBT.  For the GBT and Parkes data sets, corrections were also made to account for offsets between the clock readings from each observatory and UTC time, obtained using data from the Global Positioning System (GPS) satellites.  In the case of Nan\c{c}ay data, recorded times are derived for UTC directly from GPS, and thus no additional clock corrections were needed.

To obtain a best-fit value for dispersion measure (DM), we averaged the GBT-derived pulse profiles into five center-frequency bins (1348, 1364, 1384, 1404, and 1424 MHz).  We performed timing on this subset of the total data set using the best-fit solution derived from all telescope data, allowing DM and its derivative to vary, while holding fixed all other system parameters.  We arrived at a value for DM ($149.6258\pm0.0006$\,pc\,cm$^{-3}$) that we then held fixed for the timing analysis on the entire data set.  We have also found evidence for the existence of a dispersion measure derivative, which we include in our timing model (see Table~\ref{tab:1802_params}).  In one observation (2006 December 13), excess time delay was observed because the pulsar-Earth line-of-sight passed near the Sun, causing a temporary increase in electron column density \citep[e.g.,][]{sns+05,yhc+07}.  To account for this, we included an arbitrary time offset for this day as a parameter to be fit in the timing analysis.  

The effects of orbital motion on the pulse arrival times were taken into account using the {\sc ell1} timing model \citep{lcw+01}, due to the near-circularity of the pulsar's orbit.  This model parametrizes the eccentricity $e$ and longitude of periastron $\omega$ in terms of the two parameters $\eta \equiv e\sin\omega$ and $\kappa \equiv e\cos\omega$, which are used in the timing fit.  It also replaces the time of periastron passage $T_0$ with the time of ascending node $T_\mathrm{asc}$ as the reference epoch to be fit in the timing model.

\begin{figure}[t]
\epsscale{1.}
\plotone{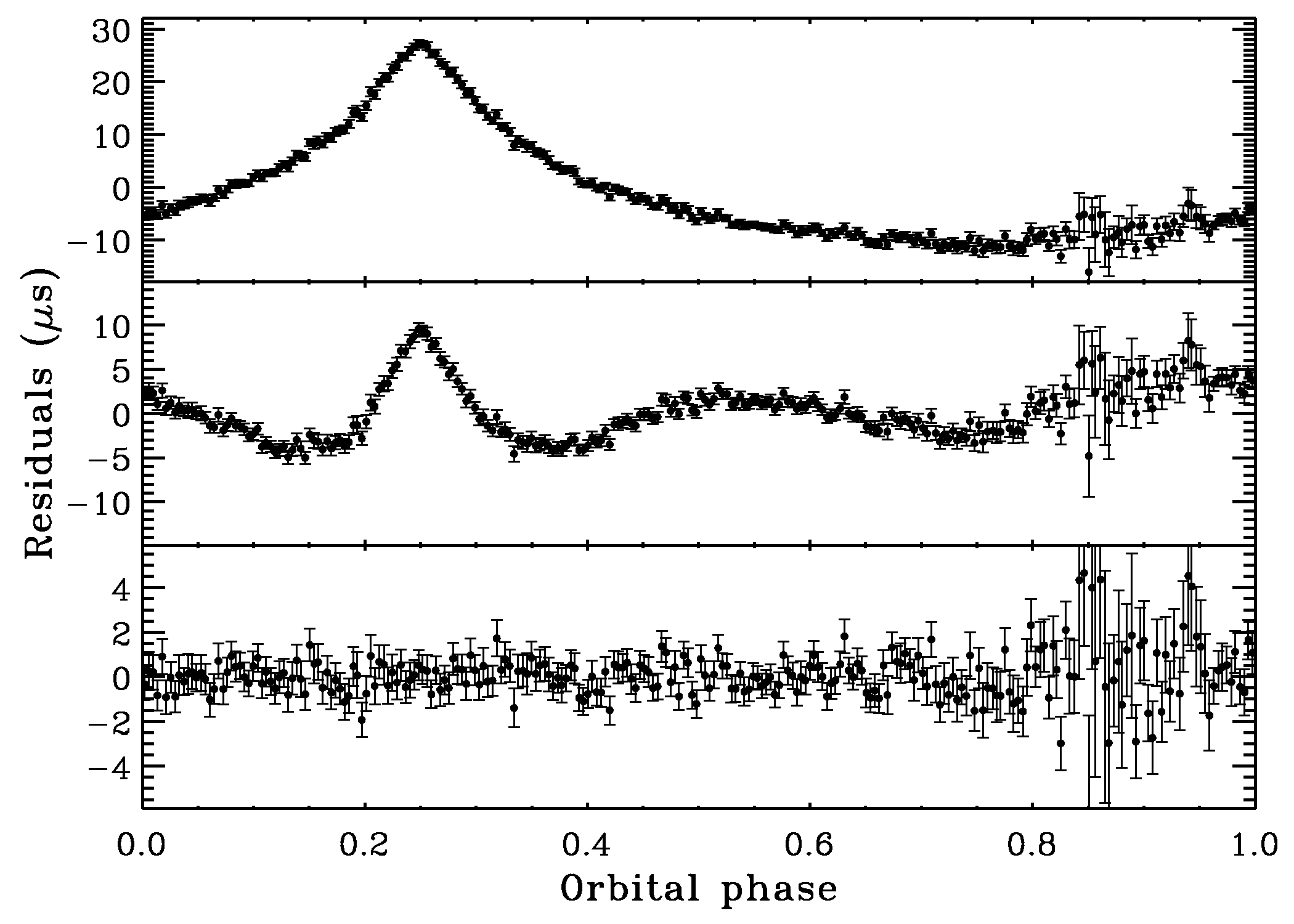}
\caption{GBT-derived timing residuals for the PSR~J1802$-$2124 system, plotted against orbital phase relative to ascending node passage. For clarity, we have averaged the residuals into 256 orbital phase bins. {\it Top}: The full effect of Shapiro delay.  Here, the Shapiro delay $r$ and $s$ parameters were excluded from the fit, with the best-fit orbital and other parameters held fixed. {\it Middle}:  Once again, the Shapiro delay terms were left out of the fit, but in this case the Keplerian orbital parameters were left to vary as free parameters.  Some of the Shapiro delay signal is absorbed into these parameters.  However, the effect of Shapiro delay is still very evident in these residuals. {\it Bottom}: All parameters, including Shapiro delay, were included in the timing model fit.\label{fig:shapiro_resids}}
\end{figure}

In addition to the basic Keplerian parameters, we fitted for the Shapiro delay of the pulsed emission as it traversed the gravitational potential well of the companion star.  This effect is described in the timing model in terms of the so-called ``range''($r$) and ``shape''($s$) parameters; the delay in the pulse arrival times for small-eccentricity orbits is given by: 
\begin{equation}
\Delta t = -2r\ln\left\{1-s\sin\left[\frac{2\pi}{\pb}(t-T_{\mathrm{asc}})\right]\right\},
\end{equation}
where $t$ is the pulse TOA.  Unless the orbit is close to edge-on, the Shapiro delay cannot be disentangled from the arrival time delay due to orbital motion \citep[see][appendix]{lcw+01}.  Figure \ref{fig:shapiro_resids} shows the timing residuals over orbital phase, resulting from various fits to the TOAs.  The effect of Shapiro delay is still very evident when fitting for the Keplerian orbital parameters, which absorb some, but not all, of the Shapiro delay signal.  Once the $r$ and $s$ parameters are measured, they can be converted into the companion mass $m_2$ and inclination angle $i$.  This is done through the following relations \citep{dd86}: 
\begin{eqnarray}
r & = & \frac{G\msun m_2}{c^3} \\
s & = & \sin i {\rm .}
\end{eqnarray}
where the relation for $r$ assumes that general relativity (GR) is the correct description of gravity, with $m_2$ expressed in solar masses.

\begin{figure}[t]
\epsscale{1.}
\plotone{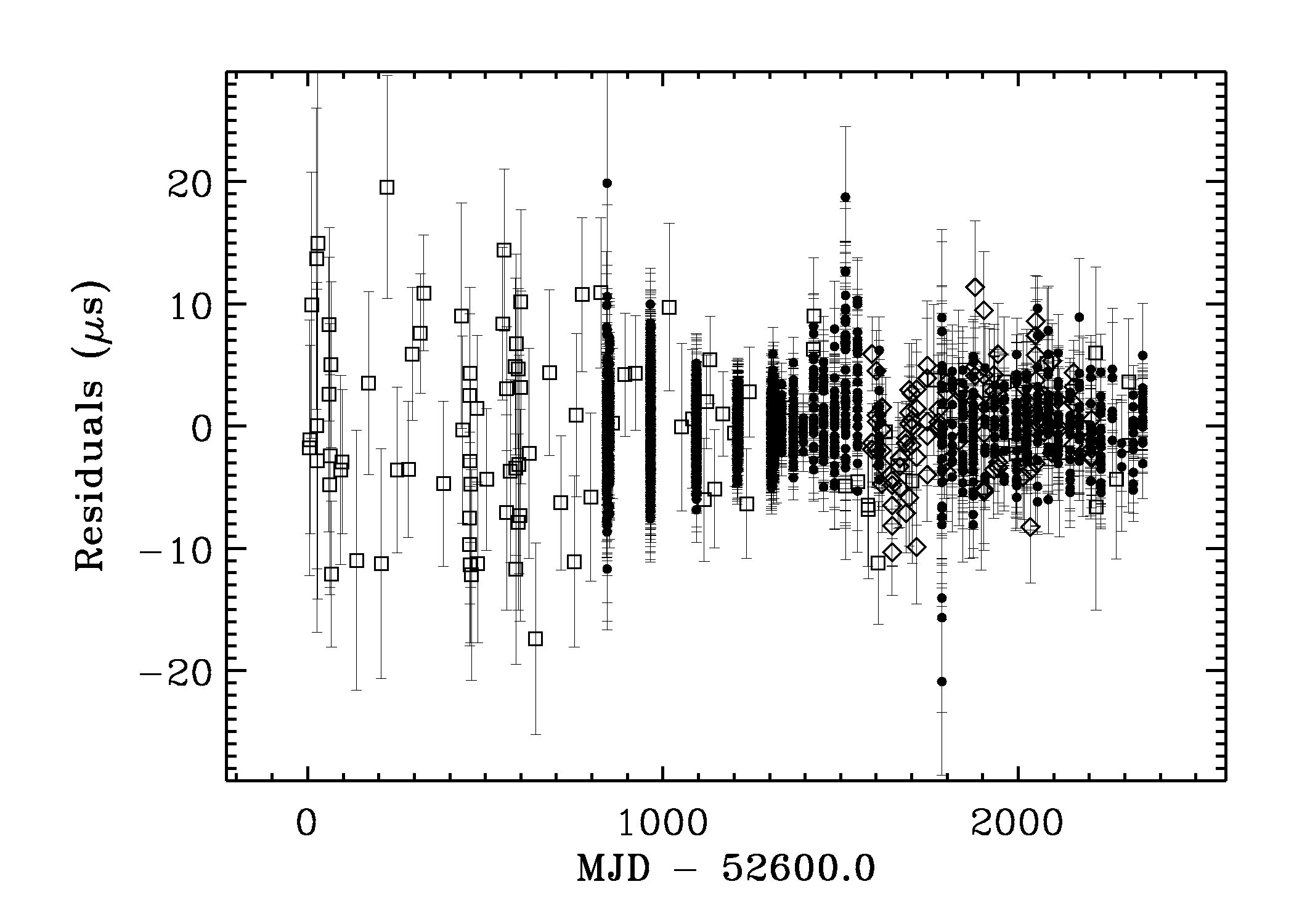}
\caption{Timing residuals for the PSR~J1802$-$2124 system plotted as a function of time (Modified Julian Date).  Parkes, GBT, and Nan\c{c}ay data are represented by open squares, filled circles, and open diamonds, respectively.\label{fig:1802_mjd_resids}}
\end{figure}

Figure \ref{fig:1802_mjd_resids} shows the timing residuals from each instrument over time.  In obtaining a best-fit model using these values, the scatter in the resulting residuals, while very close to having a random Gaussian distribution about zero, was larger than most of the errors on the individual data points, which were derived from the template profile cross-correlations.  This resulted in an overall value of $\chi^2$ per degree of freedom $\nu$ that is greater than one ($\chi^2/\nu = 1.18$ for Parkes and GBT data, and $1.47$ for Nan\c{c}ay).  This was almost certainly due to an underestimation of the TOA uncertainties that resulted from the profile cross-correlation process, or from lower-quality profiles that arose because of signal contamination by radio frequency interference, or perhaps coarse signal quantization as the data were sampled.  To compensate, we have calculated an amount to add in quadrature to
the original uncertainties in the TOAs, so that $\chi^2/\nu = 1.0$ for each telescope data set.  The GBT TOAs had errors that required very little correction, and dominated the data set.  We thus, unless otherwise noted, report the uncertainties directly output by {\tt tempo} as $68\%$ confidence limits on the fit parameters, shown in Table~\ref{tab:1802_params}.  The individual weighted rms values of the post-fit timing residuals for this pulsar are $6.1\,\mu$s from Parkes data, $2.2\,\mu$s from the GBT data, and $3.6\,\mu$s from the Nan\c{c}ay data.  The combined value for the weighted rms of the residuals is $2.3\,\mu$s.  
Another compensation method that is used to arrive at $\chi^2/\nu = 1.0$ involves the multiplication of the TOA uncertainties by calculated scaling factors. In doing this, we find a weighted rms of the post-fit residuals of $2.2\,\mu$s, an approximately $5\%$ lower value.  However, we report the slightly more conservative parameter measurements found by using the former method; these can be found in Table~\ref{tab:1802_params}.

It should be noted that we obtain significant measurements of several spin frequency derivatives from our timing analysis (see Table~\ref{tab:1802_params}).  This may be attributed to intrinsic pulsar timing instability, though this is not typically seen in recycled pulsars---including those few with similar rotation periods and surface magnetic fields to \psr{1802} (e.g., PSRs J0900$-$3144 and J1804$-$2717; see \citet{bjd+06} and \citet{hlk+04}, respectively)---with some exceptions \citep[e.g.,][]{ktr94,mte97,vbc+09}.  It may also be due to unmodeled effects from the interstellar medium along the direction of \psr{1802}, or from intervening material within the Solar System that is detectable as a result of the low ecliptic latitude of this pulsar.  While fitting for these higher-order frequency derivatives \citep[so-called ``polynomial whitening''; see, e.g.,][]{lk05} may affect the measurement of astromteric parameters such as position and proper motion (reflected in Table~\ref{tab:1802_params} by quoting $2\sigma$ uncertainties), this long-term trend did not have a significant effect on our measurements of the orbital parameters, which is our focus in this work.
 
\begin{deluxetable*}{lc}
\tablecaption{Measured and derived parameters for PSR J1802$-$2124\label{tab:1802_params}}
\tablecolumns{2}
\tablehead{
\multicolumn{2}{c}{Measured parameters}
}
\startdata
 Data span (MJD)\dotfill &     
    $52605.2-54950.4$ \\
 Right Ascension, $\alpha$ (J2000)\dotfill &     
    $18\fhour02\fmin05\fsec335576(9)\ (2\sigma)$ \\
 Declination, $\delta$ (J2000)\dotfill &     
    $-21\fdeg24^\prime03\farcs649(3)\ (2\sigma)$ \\
 Proper motion in $\alpha$, $\mu_{\alpha}$ (mas yr$^{-1}$)\dotfill &
    $-0.85(10)\ (2\sigma)$  \\  
 Proper motion in $\delta$, $\left|\mu_{\delta}\right|$ (mas yr$^{-1}$)\dotfill &
    $< 4.8\ (2\sigma)$  \\  
 Rotation frequency, $\nu$ (s$^{-1}$)\dotfill &     
    $79.066424229950(2)$  \\
 Rotation frequency derivative, $\dot{\nu}$ ($10^{-16}\,$s$^{-2}$)\dotfill & 
    $-4.5360(16)$  \\
 Reference Epoch (MJD)\dotfill &     
    $53453.0$ \\
 Dispersion measure, DM (pc\,cm$^{-3}$)\dotfill  &     
    $149.6258(6)$ \\
 Derivative of dispersion measure, $\dot{\mathrm{DM}}$ ($10^{-5}\,$pc\,cm$^{-3}$\,yr$^{-1}$)\ldots  &     
    $-9(3)$ \\
 Projected semimajor axis, $x$ (lt-s)\dotfill &     
    $3.7188533(5)$  \\
 Orbital period, $P_b$ (days)\dotfill &     
     $0.698889243381(5)$  \\
 Epoch of ascending node passage, $T_{\mathrm{asc}}$ (MJD)\dotfill &     
    $53452.633290841(4)$  \\
 $\eta \equiv e\sin{\omega}$\dotfill &
    $0.00000086(9)$  \\ 
 $\kappa \equiv e\cos{\omega}$\dotfill &
    $0.00000232(4)$ \\
 Cosine of inclination angle\tablenotemark{a}, $\left|\cos{i}\right|$\dotfill &
    $0.176(11)$\\
 Companion mass\tablenotemark{a}, $m_2\ (\msun)$\dotfill & 
    $0.78(4)$ \\
\cutinhead{High-order rotation frequency derivatives}
 Second frequency derivative ($10^{-25}\,$s$^{-3}$)\dotfill  & 
    $-1.04(7) $   \\      
 Third frequency derivative ($10^{-32}\,$s$^{-4}$)\dotfill  & 
    $-1.15(11) $   \\      
 Fourth frequency derivative ($10^{-40}\,$s$^{-5}$)\dotfill  & 
    $9.2(5) $   \\      
 Fifth frequency derivative ($10^{-47}\,$s$^{-6}$)\dotfill  & 
    $2.4(4) $   \\      
 Sixth frequency derivative ($10^{-54}\,$s$^{-7}$)\dotfill  & 
    $-4.7(4) $   \\      
 Seventh frequency derivative ($10^{-61}\,$s$^{-8}$)\dotfill  & 
    $1.45(11) $   \\      
\cutinhead{Derived parameters}
 Rotation period, $P$ (s)\dotfill &
    $0.0126475935865227(3)$  \\
 Rotation period derivative, $\dot{P}$ ($10^{-20}\,$s\,s$^{-1}$)\dotfill &
    $7.256(3)$  \\
 Distance to pulsar\tablenotemark{b}, $d$ (kpc)\dotfill &
   $2.94$  \\
 Characteristic age, $\tau_c$ (Gyr)\dotfill &
   $2.76$ \\
 Surface magnetic field, $B_{\mathrm{surf}}$ ($10^8\,$G)\dotfill &
   $9.7$  \\
 Eccentricity, $e$ ($\times 10^{-6}$)\dotfill &     
   $2.48(5)$   \\
 Longitude of periastron, $\omega$ ($^\circ$)\dotfill &
    $20(2)$  \\
 Epoch of periastron passage, $T_0$ (MJD)\dotfill &
    $53452.673(4)$ \\
 Orbital inclination\tablenotemark{a}, $i$\ ($^\circ$)\dotfill &     
    $79.9$ or $100.1(6)$   \\
 Mass function, $f\ (\msun)$\dotfill &     
    $0.11305589(5)$   \\
 Pulsar mass\tablenotemark{a}, $m_1\ (\msun)$\dotfill & 
    $1.24(11)$  
\enddata
\tablecomments{Unless otherwise noted, parentheses indicate the $1\sigma$ uncertainties on the last digit (or last two digits, if two digits are given).}\\
\tablenotetext{a}{Reported mass and $\left|\cos{i}\right|$ values represent the median value from the respective probability density functions.}
\tablenotetext{b}{The distance is estimated using the NE2001 Galactic free electron distribution model \citep{cl02}.}
\end{deluxetable*}

\section{Results}

\begin{center}
\begin{figure}[t]
\epsscale{1.}
\plotone{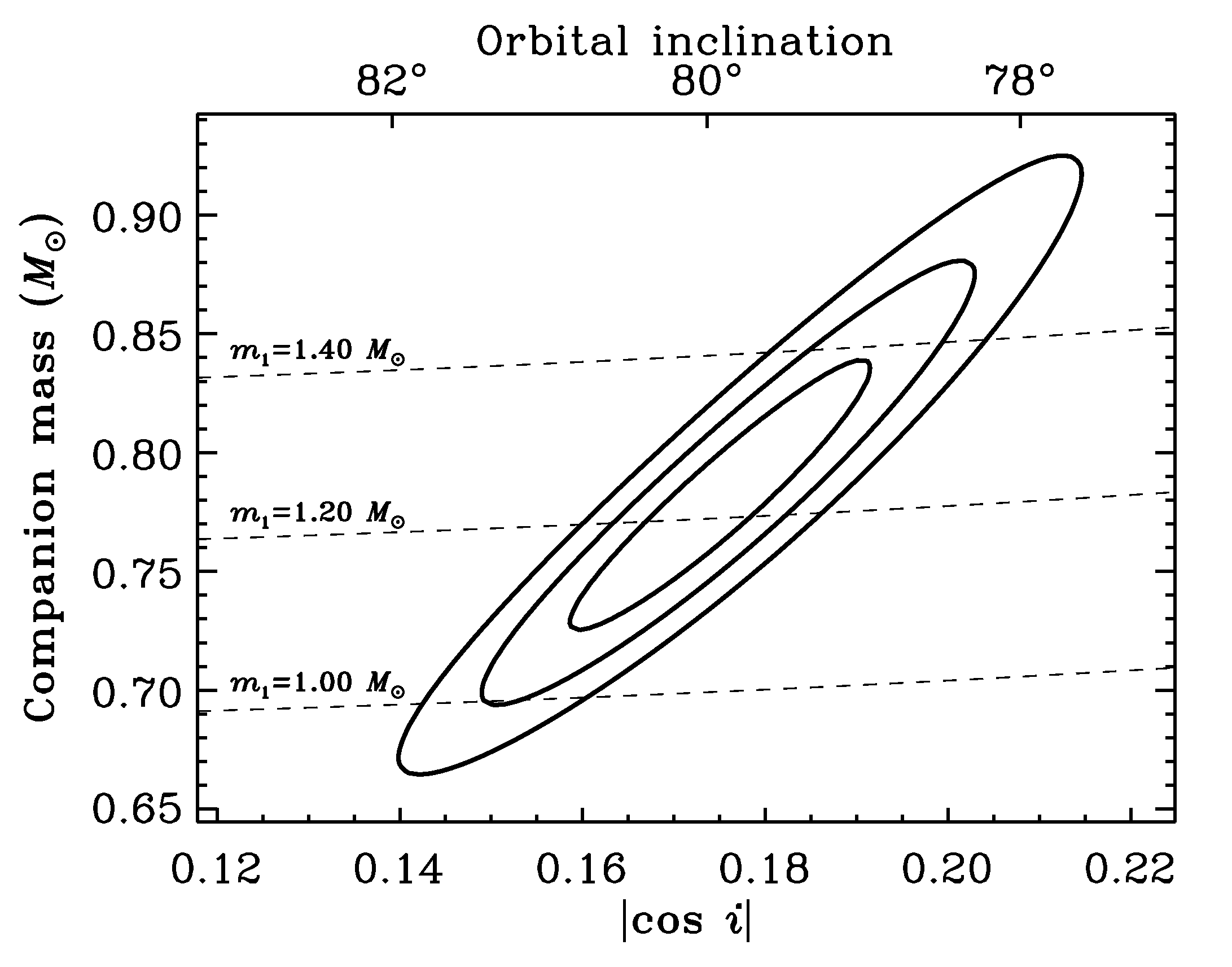}
\caption{68\% (center), 95\%, and 99.7\% confidence contours for PSR~J1802$-$2124 in orbital inclination--companion mass space. Curves of constant pulsar mass are plotted over the contours.\label{fig:1802_contour}}
\end{figure}
\end{center}

In order to ensure that the measured system masses represent the best model fit, we probed the $\chi^2$ over a fine grid of values, evenly distributed in $\left|\cos{i}\right|-m_2$ space, and allowing all other timing parameters to vary. (We use the absolute value of $\cos{i}$ since we cannot distinguish $i < 90\degrees$ from $i > 90\degrees$.)  The resulting confidence contours mapped by the $\chi^2$ values within our grid are shown in Figure \ref{fig:1802_contour}.  Overlaid are curves of constant pulsar mass $m_1$, which is calculated through the Keplerian mass function, given by:
\begin{equation}
f(m_1,m_2,i) \equiv \frac{(m_2\sin i)^3}{(m_1 + m_2)^2} = x^3\left(\frac{2\pi}{\pb}\right)^2\left(\frac{1}{\tsun}\right){\rm ,}
\end{equation}
where $x\equiv a\sin{i}/c$ is the projected semi-major axis of the pulsar's orbit, and $\tsun \equiv G\msun/c^3 = 4.925490947\times 10^{-6}\,$s is the mass of the Sun in units of seconds.  The mass function $f$ and system masses $m_1$ and $m_2$ are expressed in units of solar mass.  

The most probable pulsar mass, companion mass, and $\left|\cos i\right|$ were found by calculating their respective marginalized probability density functions (PDFs).  Details of the method can be found in \citet[][appendix]{sna+02}.  The intervals representing the $68\%$ uncertainty in these quantities were calculated by determining the parameter values to either side of which the tails of each PDF cover $16\%$ of the total area under the respective functions.  We find that the best-fit median pulsar and companion masses are $1.24\pm0.11\msun$ and $0.78\pm0.04\msun$, respectively.  This represents the second-most precise timing measurement to date of a pulsar in a NS-WD binary, after PSR~J1909$-$3744 \citep{jhb+05}.  We note that the weighted rms of the post-fit timing residuals varies by less than $10\%$ when the system masses and inclination angle are fixed at $\pm 3\sigma$ from their best-fit values.

We have also measured the right-ascension component of the system's proper motion to be $\mu_{\alpha}=-0.85\pm0.10\,\mathrm{mas\,yr}^{-1}$ ($95\%$ uncertainty).  The pulsar's small ecliptic latitude has made it difficult to measure its proper motion in declination with the current data set.  We thus quote a $95\%$ upper limit of $4.8\,\mathrm{mas\,yr}^{-1}$ for this quantity.  However, we can calculate a one-dimensional space velocity based on our measurement of $\mu_{\alpha}$ and estimated distance of 2.94 kpc, based on the NE2001 Galactic free electron distribution model \citep{cl02}, given the measured pulsar DM.  We find that the pulsar velocity in the right ascension direction $v_\alpha = $12 km\,s$^{-1}$, suggesting a relatively low velocity compared with those of other millisecond and binary pulsars \citep[see, e.g.,][]{tsb+99,lkn+06}.  See \S 7 for further discussion of the pulsar velocity.

\section{Evolution of the PSR~J1802$-$2124 system}

The timing results from observations of PSR~J1802$-$2124 over the past six and a half years show that it is in a relatively compact binary system with a massive WD.  It is also a light pulsar; along with others such as \psr{1141} \citep{bokh03} and \psr{1713} \citep{sns+05} for example, it is among the least massive known NSs with WD companions.  This mass measurement represents the first made for what we refer to as \emph{short orbital-period} IMBPs ($\pb < 3\,$days; see Table~\ref{tab:imbp}). 

We find that several NS-WD binary formation scenarios cannot explain the observed parameters of the \psr{1802} system.  The usual LMBP mass transfer scenario, which would invoke an extended, stable period of accretion of matter onto the NS surface, is difficult to reconcile with our measurements of the pulsar and WD companion masses; it also seems to be incompatible with the measured pulsar rotation period, which is significantly longer than those typically found in LMBP systems.  The highly super-Eddington accretion scenario outlined earlier \citep{tvs00} also does not appear to be able to produce the \psr{1802} system---while the companion is likely to be a CO WD ($0.4 \lesssim m_2 \lesssim 0.9\,\msun$), the 16.8-hour orbital period is significantly less than the $\sim3$-day minimum period produced in this scenario.  The double-He core progenitor scheme of B01, used to explain the formation of PSR~B0655+64, assumes that the progenitors of the NS and WD have similar masses.  However, the NS mass in the \psr{1802} system differs significantly from that of the WD companion.  This probably indicates a corresponding disparity in mass for their progenitors, making this theory difficult to apply in this case.  This leads us to believe that the most viable formation scenario for the \psr{1802} system is that it had survived a phase of CE evolution.  This is supported by the compactness of the orbit, the large WD mass, and moderately slow spin rate.  Our mass measurements indicate that the pulsar probably had little time to accrete matter from the companion star before the envelope was ejected \citep[$\sim10^3$ years; e.g.,][]{tv06corr}, as its mass is similar to those of recycled pulsars in DNS systems \citep[e.g.,][]{nsk08, sta08}

\section{Common-envelope survival rate of IMBPs}

Although we believe the evolution of the PSR~J1802$-$2124 system to be generally understood, the precise histories of this system and other short orbital-period IMBPs can only be elucidated through detailed binary stellar evolution simulations, which are well beyond the scope of this paper.  We now briefly discuss in broad terms several potential contributions to the discordance between the theorized and observed population of IMBPs.

\subsection{Observational clues}

Using the population synthesis results of \citet{py98}, as well as evolution analysis by \citet{bb98b},  B01 argue that, within a factor of two, the ratio of the birthrate of short orbital-period IMBPs (i.e., those which have passed through a CE phase in their evolution) to that for the young eccentric NS-massive WD binaries is expected to be $\sim1$.  Only two of the latter class have been observed \citep{std85,vk99,klm+00}.  B01 also argue that due to their longer visible lifetimes (due to their approximately two orders of magnitude weaker surface magnetic fields), we should observe a factor of $\sim50-100$ more short orbital-period IMBPs, while only four or five have so far been seen.  According to this argument,  hypercritical accretion-induced collapse of most CE-embedded NSs into black holes is responsible for this observational discrepancy.  

We also note that through evolutionary model analysis, \citet{bkb02} find that roughly two-thirds of NS systems can survive hypercritical accretion for a maximum NS mass greater than $2\msun$.  However, we find a relatively low mass for \psr{1802}, indicating that this pulsar (as well as those in the precisely-measured pulsars in DNS systems) has not accreted a significant amount of matter, regardless of the specific formation mechanism undergone by these systems.

The wide range of observed 1400-MHz luminosities for both short orbital-period IMBPs and eccentric NS-WD systems make a rigorous comparison of the expected and observed populations difficult without exploring an array of population synthesis models \citep[e.g.,][]{kklw04,kkl+05} or considering the various survey selection effects, some of which we now outline.

\begin{center}
\begin{deluxetable*}{lcccccccc}
\tablecaption{Known short orbital-period intermediate-mass binary pulsars and eccentric neutron star-white dwarf systems\label{tab:imbp}}
\tablecolumns{9}
\tablehead{
  \colhead{PSR} & 
  \colhead{Rotation} &
  \colhead{Orbital} &
  \colhead{Eccentricity} &
  \colhead{Surface} &
  \colhead{Flux density} &
  \colhead{Distance\tablenotemark{a}} &
  \colhead{Luminosity density} &
  \colhead{References}  \\
  \colhead{} & 
  \colhead{period} &
  \colhead{period} &
  \colhead{} &
  \colhead{magnetic field} &
  \colhead{(1400 MHz)} &
  \colhead{(kpc)} &
  \colhead{(1400 MHz)} &
  \colhead{} \\
  \colhead{} & 
  \colhead{(ms)} &
  \colhead{(days)} &
  \colhead{} &
  \colhead{($10^{10}\,$G)} &
  \colhead{(mJy)} &
  \colhead{} &
  \colhead{(${\rm mJy\, kpc}^2$)} &
  \colhead{}
}
\startdata
\multicolumn{9}{c}{Short orbital-period IMBPs ($P_{\mathrm{b}} < 3\,\mbox{days}$)}\\[0.2pc]
\hline\\[-0.2pc]
 B0655+64  &         
  195.7             &  1.03\phn    & 0.0000075     &   $1.17$\phn\phn
  &   0.3\phn   &  0.49  &   \phn\phn\phn\phd0.9  &  1,2
    \\
 J1232$-$6501  &        
  \phn88.2          &  1.86\phn    & 0.00011\phn\phn     &   $0.856$\phn
  &   0.34  &  6.2\phn  &   160   &  3,4
    \\
 J1435$-$6100  &        
  \phn\phn\phn9.34  &  1.35\phn    & 0.0000105     &   $0.0484$
  &   0.25  &  2.2\phn  &   \phn15   &  3,4 
    \\
 J1757$-$5322  &     
  \phn\phn\phn8.87  &  0.453       & 0.0000040     &   $0.0489$
  &   2.3\phn   &  0.96  &  \phn27   &  5,6
    \\
 J1802$-$2124  &     
  \phn12.6          &  0.699       & 0.0000025      &   $0.0966$
  &   0.77  &  2.9\phn  &   \phn84   &  7, this work
    \\
\cutinhead{Eccentric NS-CO WD binaries}\\[-0.4pc]
 J1141$-$6545  &     
  394\phn\phd            &  0.198         &   0.17\phn\phn\phn\phn\phn      &   132\phn\phn\phn\phn\phn\phn\phd
  &   3.3\phn   &   $>3.7$\tablenotemark{b}\phn\phn   &   $>570$\phn\phn  &  8,9 \\
  B2303+46 &     
  1066\phn\phn\phd            &  12\phn\phn\phn\phn\phd   &   0.66\phn\phn\phn\phn\phn      &   78.8\phn\phn\phn\phn
  &  \phn0.38\tablenotemark{c}   &  \phn\phn2.9740  &  \phn42   &  10
    \\[-0.4pc]
\enddata
\tablerefs{1.---\citet{jl88}; 2.---\citet{lylg95}; 3.---\citet{clm+01}; 4.---\citet{mlc+01}; 5.---\citet{eb01b}; 6---Bailes, private communication; 7.---\citet{fsk+04}; 8.---\citet{klm+00}; 9.---\citet{obv02a}; 10.---\citet{std85}.}
\tablenotetext{a}{For all pulsars, except where noted, we have used the NE2001 model \citep{cl02} to derive the distances, based on the dispersion measures of the objects.}
\tablenotetext{b}{For PSR J1141$-$6545 we have used the distance published in \citet{obv02a}, derived from the neutral hydrogen absorption spectrum along the pulsar's line of sight.}
\tablenotetext{c}{PSR~B2303+46 does not have a published 1400 MHz flux density.  In this case we estimate it using published spectral index for this pulsar, from \citet{mkk+00}.}
\end{deluxetable*}
\end{center}

\subsection{Selection effects}
We present here two important selection effects against the observation of IMBPs that have been neglected by B01.  Firstly, for a given luminosity and DM, a shorter spin period will render the pulsar more prone to the observational effects of dispersive smearing.  This is because, in faster-rotating pulsars, the pulse will become smeared to a greater extent as a fraction of the spin period.  This applies here, since IMBPs show an overall spin period distribution that is substantially shorter than in the eccentric NS-WD binaries (see Table~\ref{tab:imbp}).  Indeed, \psr{1802} has a spin period-to-DM ratio that is approximately a factor of 40 smaller than, for example, \psr{1141} ($P_{\mathrm{spin}}/\mathrm{DM} \sim 0.084$ and $3.4$, respectively).  To further illustrate this point, we have calculated the observed fractional pulse width as a function of DM for two pulsars at 1400 MHz, assuming 3-MHz channels.  This is similar to the search observation setup for the Parkes Multibeam Pulsar Survey \citep[e.g.,][]{mlc+01}.  An eccentric NS-WD binary pulsar like PSR~J1141$-$6545 \citep[$P_{\rm spin}=394\,$ms;][]{klm+00} has a fractional pulse width that would never be smeared by more than $2\%$ of the pulse period, out to a DM $\gtrsim 1200\,{\rm pc\,cm^{-3}}$, near the limit of the known pulsar population; hence dispersion smearing is negligible for this class of pulsars.  In contrast, an IMBP like PSR~J1802$-$2124 ($P_{\rm spin}=12.6\,$ms) has a observed fractional pulse width that grows to more than one-sixth of a pulse period, making the pulsar relatively difficult to detect, at DM $\sim190\,{\rm pc\,cm^{-3}}$ or larger.  This implies that the eccentric young-pulsar binaries can be discovered to much larger volumes than the IMBPs.  The largest DM to which an IMBP has thus far been discovered is PSR~J1810$-$2005, at $240.2\,{\rm pc\,cm^{-3}}$ \citep{clm+01}, which has a spin period approximately three times that of \psr{1802}.

To quantify the relative detectability and survey volume due to propagation effects, we performed the following simple Monte Carlo population using the freely available {\tt psrpop} pulsar population modelling package \citep{lfl+06}. We generated model galaxies containing pulsars with periods and pulse widths identical to \psr{1802} and PSR~J1141$-$6545. Each model population is distributed in Galactocentric radius according to the best-fit distribution found by \citet{lfl+06}. To test whether there is any dependence on the dispersion above the Galactic plane, we generated  models with exponential $z$-height distributions with means in the range 300--500~pc. Since we are only interested in propagation effects on the relative detections of the two pulsars, we assign each model pulsar a radio luminosity of 100~mJy~kpc$^2$. For each population, we then record the  number of detected pulsars in the detailed model of the Parkes Multibeam Pulsar Survey described by \citet{lfl+06}.  Regardless of the assumed $z$-scale height, the ratio of detections of PSR~J1141$-$6545 to \psr{1802} is always $\sim 2.2$.  In other words, due to propagation effects alone, the effective volume of the Galaxy surveyed by the Parkes Multibeam Survey for \psr{1802}-like objects is only half that of pulsars similar to PSR~J1141$-$6545. 
It should be noted that this simulation does not take account of any beaming corrections which are likely to be significant given that longer period pulsars generally have smaller beams than their shorter period counterparts \citep[e.g.,][]{tm98}.  To quantify this, using the \citet{tm98} beaming model, we estimate that the beaming fraction of \psr{1802} is four times larger than that of \psr{1141}. When combined with the above simulations, this would imply that for populations of comparable sizes, one might expect only half as many objects that are \psr{1141}-like compared to those that are similar to \psr{1802}.
Future surveys that use narrower-channel instruments should expect to find a larger number of higher-DM IMBPs compared to eccentric NS-WD binaries.

We also note that pulsar acceleration would cause further signal spread in the Fourier search domain, to a larger extent for IMBPs than for young pulsars in similar orbits \citep[e.g.,][]{hrs+07}.  Although difficult to quantify, this is an important additional selection effect against discovery of IMBPs relative to the eccentric NS-WD binaries.  

While the discrepancy in the observed ratio of these two system types pointed out by B01 may still be supported by the available data on IMBPs, we emphasize again that a more precise estimation of observable numbers of short orbital-period IMBPs will only come with further population synthesis studies, as well as accounting carefully for survey selection effects.

\section{Future measurements and studies}

A more precise measurement of the full proper motion will help us to constrain the space velocity of this system.  This will further our understanding of IMBP formation history.  For example, it would allow us to test the apparently low scale height of IMBPs compared to LMBPs, presumably due to the larger combined mass of IMBP progenitor systems, as suggested by \citet{clm+01}.  The low transverse velocity that is hinted at by the measured proper motion in the right ascension direction supports this theory.  Within five years, we expect to obtain a significant measurement of proper motion in declination.  This will enable us to calculate a more reliable value for the transverse velocity of \psr{1802}.

Based on the measured orbital parameters, the GR prediction for orbital decay $\pbdot^{\mathrm{GR}} = -3.1\times10^{-14}$ s\,s$^{-1}$ for the \psr{1802} system, and we expect to significantly measure $\pbdot$ in the near future.  This has only been achieved in four NS-WD systems: in the case of PSR~J1141$-$6545 \citep{klm+00a,bokh03,bbv08} and PSR~J0751+1807 \citep{nss+05,nsk08}, the $\pbdot$ intrinsic to the system is measured; in the PSR~J1012+5307 system \citep{lwj+09}, the intrinsic $\pbdot$ is not yet significantly determined, however the contribution to the measured value from kinematic effects is of the same order as the quadrupolar GR prediction; and in PSR~J0437$-$4715 \citep{vbs+08}, the measured $\pbdot$ is attributed predominantly to contamination by apparent acceleration due to the system space velocity, as the measured $\pbdot$ is four orders of magnitude greater than the value predicted by GR for that system.

For \psr{1802}, we anticipate an intrinsic $\pbdot$ measurement which, when combined with the Shapiro delay measurements, will overdetermine the system mass values and allow for a check on the relativistic analysis of the kinematics of the system.

The difference in self-gravities between \psr{1802} and its WD companion is predicted to cause a deviation in $\pbdot$ from the quadrupolar GR prediction according to some scalar-tensor theories of gravity \citep[e.g.,][]{esp05corr}.  A limit to this departure from the GR value could be used to constrain the existence of dipolar radiation from this system, or to set a limit to the variation of the gravitational constant, $\dot{G}$ \citep{nor90,dt91,lwj+09}.  It remains to be seen, however, to what extent the kinematic corrections \citep{dt91,nt95} may affect our ability to use this measurement for any of these purposes.

\section{Conclusions}
In this paper we have described the observations and timing analysis of \psr{1802}.  These have provided updated system parameters and precise measurements of the pulsar and companion WD masses, which we find to be $1.24\pm0.11\msun$ and $0.78\pm0.04\msun$, respectively.  In particular, this determination was made possible by the detection of Shapiro delay on the pulse arrival times.  The result is of particular significance for this class of pulsar, since it is only the second such mass measurement for an IMBP system, and the first for a short orbital-period IMBP ($\pb \lesssim 3\,$days).  The mass measurements of the \psr{1802} system highlight the dependence of the final system configuration on the specific mass-transfer history and particular evolution of the system in question.  The similarity between the properties of this system to those of recycled DNS pulsars hints that the evolutionary paths of these two system types may be analogous, probably involving a CE/inspiral phase.  

It is clear that to arrive at a definitive picture of IMBP evolution, and more generally, the evolution of the many observed binary system types, we must discover more systems with measurable masses, and several recent and future surveys \citep[e.g.,][]{vls04,cfl+06,lsf+06} are expected to find many such systems for study.

\acknowledgements
The authors would like to thank the referee for the constructive comments.  We thank Caltech, Swinburne University, and NRL pulsar groups for use of the CGSR2 cluster at Green Bank.  We wish to thank M.~Bailes for providing flux density information on those pulsars noted in Table~\ref{tab:imbp}.  We would also like to thank J.~Verbiest, M.~Bailes, and B.~Jacoby for helping our understanding the GBT clock history, and to J.~Verbiest for useful discussions about timing stability.  Thanks as well to W.~van Straten for his help with the \texttt{psrchive} software.  RDF was partially funded by a UBC UGF award.   IHS  held an NSERC UFA during part of this work, and also acknowledges sabbatical support from the ATNF Distinguished Visitor program and from the Swinburne University of Technology Visiting Distinguished Researcher Scheme.  Pulsar research at UBC is supported by an NSERC Discovery Grant.  GASP is funded by an NSERC RTI-1 grant to IHS and by US NSF grants to DCB (AST 9987278, 0206044) and DJN (AST 0647820).  MAM and DRL are supported by WVEPSCoR via a Research Challenge Grant.  PBD is a Jansky Fellow of the National Radio Astronomy Observatory.  The Parkes radio telescope is part of the Australia Telescope which is funded by the Commonwealth of Australia for operation as a National Facility managed by CSIRO. The National Radio Astronomy Observatory is a facility of the U.S. National Science Foundation operated under cooperative agreement by Associated Universities, Inc.  The Nan\c{c}ay radio telescope is part of the Paris Observatory, associated with the Centre National de la Recherche Scientifique (CNRS), and partially supported by the R\'{e}gion Centre in France.



\end{document}